\documentclass[printer]{tPHM2e}
\usepackage{rotating}
\usepackage{amsmath,amssymb,graphicx,psfrag}

\newcommand{\scs}{\scriptscriptstyle}

\begin{document}

\title{On the origin of the Boson peak in globular proteins}

\author{Stefano Ciliberti\thanks{$^\ast$Corresponding author. Email:
        Stefano.Ciliberti@roma1.infn.it. Now at Science \& Finance,
        Capital Fund Management, 6 Blvd Haussmann, 75009 Paris,
        France.}$^\ast$\\\vspace{6pt} CNRS; Univ. Paris Sud, UMR8626,
        LPTMS, ORSAY CEDEX, F-91405\\\vspace{8pt} Paolo De Los Rios,
        Francesco Piazza\\\vspace{6pt} LBS, Institute of Theoretical
        Physics - BSP-720, EPFL, CH-1015 Lausanne, Switzerland}


\maketitle

\begin{abstract}
  We study the Boson Peak phenomenology experimentally observed in globular
  proteins by means of elastic network models. These models are suitable for
  an analytic treatment in the framework of Euclidean Random Matrix theory,
  whose predictions can be numerically tested on real proteins
  structures. We find that the emergence of the Boson Peak is strictly
  related to an intrinsic mechanical instability of the protein, in close
  similarity to what is thought to happen in glasses. The biological
  implications of this conclusion are also discussed by focusing on a
  representative case study.
\end{abstract}


\section{Introduction \label{sect:intro}}

The analogy between proteins and structural glasses has a long story and
solid foundations. Among the different physical properties shared by the two
classes of systems one can mention: (i) the anomalous specific
heat~\cite{green94}; (ii) the slow energy relaxation
processes~\cite{frauenfelder88, xie02, piazza05}; (iii) the existence of a
dynamical transition as witnessed by the temperature dependence of the the
atomic mean squared displacements ~\cite{doster89, iben89}; (iv) the excess
of low frequency modes (the Boson Peak) in dynamical spectra obtained by
inelastic neutron scattering experiments~\cite{orecchini01, leyser99,
doster90}.  Although a somewhat generic explanation of the above
experimental facts can be given in terms of a rugged energy
landscape~\cite{ansari85, onuchic97, frauenfelder98}, a comprehensive
explanation supported by solvable models is still missing.  Incidentally, it
is no surprise that proteins have been proposed as the paradigm of
complexity in biological systems~\cite{frauenfelder02}.

Here we focus on one of the glass--like features of proteins, the Boson Peak
(BP). The latter is formally defined as the low frequency peak in the
function $g(\omega)/g_{D}(\omega)$, where $g(\omega)$ is the vibrational
density of states (DOS) as measured experimentally, and $g_{D}(\omega)
\propto \omega^2$ is the Debye behavior in the $\omega \to 0$ limit. A peak
in this function would then represent an excess of low vibrational modes
with respect to a perfect harmonic crystal. The very origin of these new
modes, related to the amorphous nature of the low energy configurations, has
been the subject of a long debate in the recent years and yet there is no
general agreement. As for structural glasses, people have proposed
``soft-potential'' theories including strong
anharmonicities~\cite{buchenau92, karpov83}, harmonic lattices with random
springs~\cite{schirmacher98, taraskin01}, extensions of standard mode
coupling theory to the ``frozen'' state~\cite{goetze00, voigtmann02},
and topologically disordered models~\cite{grigera01, ciliberti03}. Recently, the
possibility that the BP is a universal feature of weakly connected systems
has been suggested~\cite{wyart05}.

In this paper we will take the point of view that proteins can be considered 
to some extent as ``random'' structures, akin to those of structural glasses.
In a sense, this constitutes our working hypothesis to be tested \textsl{a
posteriori}. However, the rationale behind it can be identified in the empirical
observation of the large scale structural properties of proteins. For instance,
the pair correlation function of proteins at
distances larger than $\approx 10\div 15$ \AA \ can be hardly distinguished from
that of a set of points randomly distributed in an equivalent volume~\cite{noiPRL}. 
As a consequence,  the ``random'' approximation is expected to work
reasonably well for studying the behaviour of low frequency modes.

The rest of the paper is organized as follows. In section \ref{sect:model}
we introduce the elastic network model of protein fluctuations and we
briefly justify its usage in this context. In section \ref{sect:ermt} we
discuss the harmonic approximation and the theoretical framework of the
Euclidean Random Matrix (ERM) theory, whose main findings are also briefly
summarized as a reference for the interpretation of our numerical
results. In section \ref{sect:results} we report our results on the BP and
on the role of the critical parameter, as well as a study of the
relationship between functional modes and proteins' stability. Our comments
and suggestions for further experimental studies are discussed in the last
section.

\section{Model building: Elastic Networks \label{sect:model}}

\begin{figure}
\begin{center}
\includegraphics[angle = 90, width=16 truecm,clip]{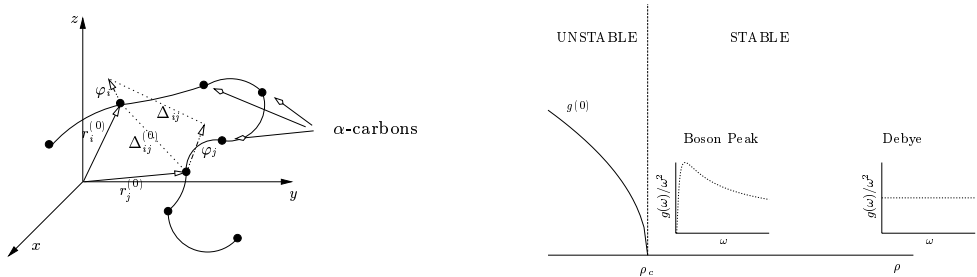}
\end{center}
\caption{Left: Schematic cartoon of an Elastic Network Model. The protein is
  seen as a polymer-like chain coarse-grained at the amino-acid level with
  elastic pairwise interactions. Right:  Pictorial representation of the
  phase diagram predicted by the ERM theory.}
\label{fig:en}
\end{figure}

The introduction of coarse-grained elastic network models in the context of
protein dynamics is quite recent~\cite{Tirion}. In fact, as a direct
consequence of the known complexity of proteins' energy landscapes, all
microscopic details are usually deemed essential to characterize their
functional dynamics. As a result, people have long failed to realize that
most features of the large- and medium-scale dynamics of proteins close to
their native state, i.e. those related to biological function, can be
successfully reproduced by simple harmonic interactions between
amino-acids~\cite{Bahar,Haliloglu,hinsen,yhs}.

In our case, we aim at describing a universal feature of globular proteins
at low frequency. Hence, the class of elastic network models appears {\em a
fortiori} the correct choice to capture the main aspects of the BP
phenomenology. In this spirit, the fine structural detail of microscopic
interactions will be completely neglected, as well as the information
contained in the amino-acid sequence. We thus coarse-grain the protein
structure at the amino-acid level and replace each residue by a single
particle whose equilibrium position coincides with that of the
$\alpha$-carbon as measured e.g. through X-ray crystallography. In practice,
this amounts to dealing with a polymer folded exactly like the real
protein. Each pair of particles falling within a fixed cutoff inter-distance
are assumed to interact harmonically. More precisely, the global Hamiltonian
of the system is (see the left panel of Fig.~\ref{fig:en} for the
notations):
\begin{equation}
  \mathcal{H}[\{{\bf r}_i\};\{{\bf r}^{(0)}_i\}]
  = \sum_{i,j} K (\Delta^{(0)}_{ij})
  \big( |{\bf \Delta_{ij}}| - |{\bf \Delta^{(0)}_{ij}}| \big)^2 \ .
  \label{eq:hamiltonian}
\end{equation}
The interaction stiffness can be also allowed to decrease smoothly with the 
pair distance as e.g. $K (\Delta^{(0)}_{ij}) = \kappa
\exp\big[-(\Delta^{(0)}_{ij}/r_c)^2\big]$ (Gaussian model), but a sharp cutoff
(i.e. stepwise) form of the type $K (\Delta^{(0)}_{ij}) = \kappa \Theta
\big[r_{c}-\Delta^{(0)}_{ij} \big]$ could also be used. 
In this paper, we will adopt the Gaussian model. 

The parameter $\kappa$ sets the physical units for force constants, and is
usually fixed by requiring the theoretical mean squared displacements of
residues to match the experimental ones as determined from X-ray spectra for
a given choice of the cutoff $r_{c}$~\cite{Bahar,Haliloglu}. In principle,
the latter should be tuned by fitting the low-frequency portion of
experimental spectra at temperatures below the dynamical transition, where
the protein fluctuates harmonically within the native minimum. However, such
studies have never appeared in the literature to the best of our
knowledge. The usual, cheaper alternative is to compare the theory with
numerical spectra obtained by modeling the protein dynamics through all-atom
force fields~\cite{hinsen}.  By doing so, one obtains $\rho_{c} \approx 3$
\AA \ in an all-atom representation, which reduces to $r_{c} \approx \langle
N_{a} \rangle^{1/3}\rho_{c} \approx 8$ \AA \ when the protein structure is
coarse-grained at the residue level. In the following, such value wil be
referred to as the optimal cutoff length scale.

\section{Harmonic approximation and Euclidean Random Matrix Theory \label{sect:ermt}}

Taking the harmonic approximation of the Hamiltonian (\ref{eq:hamiltonian})
one is led to the quadratic form $\mathcal{H} \simeq \frac 12 ( {\bf
\varphi}, \mathcal{M} {\bf \varphi})$, where the Hessian matrix
$\mathcal{M}$ depends on the $\alpha$-carbon coordinates $\{r^{(0)}_i\}$. As
we anticipated in section \ref{sect:intro}, we will investigate the
consequences of the hypothesis that these positions are random in
space. Under this assumption, the matrix $\mathcal{M}$ falls into the broad
class of Euclidean Random Matrices~\cite{mezard99}. In general terms, our
problem amounts to taking $N$ points at random in a volume $V$, assume that
they interact according to some pair potential $v(r)$ and then compute the
vibrational DOS of this \emph{topologically disordered} system in the
thermodynamic limit $N,V\to\infty$, while the density $N/V=\rho$ stays
finite. After some efforts~\cite{martin01, grigera01, ciliberti03,
grigera03}, the following regimes have been identified under rather generic
assumptions (see also the right panel of Fig.~\ref{fig:en}):
\begin{itemize}
\item
  {$\rho \gg \rho_c$}, high-density / low-temperature: the DOS is found to
  follow exactly the Debye behavior, $g(\omega) \propto \omega^2$. An
  amorphous solid behaves like a perfect crystal in the limit of very large
  density, meaning that an infinitely compact medium does not feel at all
  the presence of the disorder.
\item
  {$\rho \gtrsim \rho_c$:} a peak in {$g(\omega)/\omega^2$} emerges at
  {$\omega_{BP}$}, where $\omega_{BP}\sim (\rho-\rho_c)$, and its height is
  found to diverge like $h_{BP} \sim (\rho-\rho_c)^{-1/2}$. This peak is
  identified with the BP and is due to the fact that the system ``feels''
  the presence of a mechanical instability at $\rho_c$. The actual value of $\rho_c$
  depends on the details of the interaction.
\item
  {$\rho < \rho_c$:} The characterization of the low density phase depends
  on the spatial nature of the interactions. In a vectorial
  model~\cite{ciliberti03} the Hessian matrix has negative eigenvalues
  (i.e. imaginary frequencies). This phenomenon is usually referred to as a
  topological phase transition. The DOS at zero energy plays the role of an
  order parameter and one finds that $g(0) \sim (\rho_c-\rho)^{1/2}$.
\end{itemize}

To summarize, the system undergoes a phase transition, from a ``solid''
phase characterized by the minima of the energy surface, to a ``liquid''
phase where saddles become relevant for the high-frequency dynamics. If the
Hessian matrix is positively defined and thus no negative modes are
possible, the DOS in the liquid phase just reduces to a delta function at
zero frequency~\cite{grigera01}. It is also interesting to mention that the
mean field exponents are consistent with the numerical findings on realistic
glass-forming materials~\cite{grigera03}.

In view of applying this approach to our protein model, a control parameter
has to be identified. The most natural choice is the interaction cutoff
$r_c$. Given that the average connectivity $\langle c \rangle \sim \ $Tr$
\mathcal{M} \sim r_c^3$, $r_c$ is indeed a global measure of compactness of
the protein. Very much like the temperature, this parameter is expected to
signal the approach of an instability akin to a liquid-glass transition:  if
$r_c$ is very large, the protein is extremely rigid and one expects its DOS
to follow the Debye law. As $r_c$ decreases, the protein looses stability
and becomes an extremely flexible object, until it unfolds and eventually
melts. If this idea is correct, one should detect a trace of this
progressive structural change by looking at the vibrational DOS upon moving
$r_c$. This is the problem we address in the next section.

\section{Results\label{sect:results} }

\subsection{The Boson Peak}

\begin{figure}
\centering
\includegraphics[width=12 truecm]{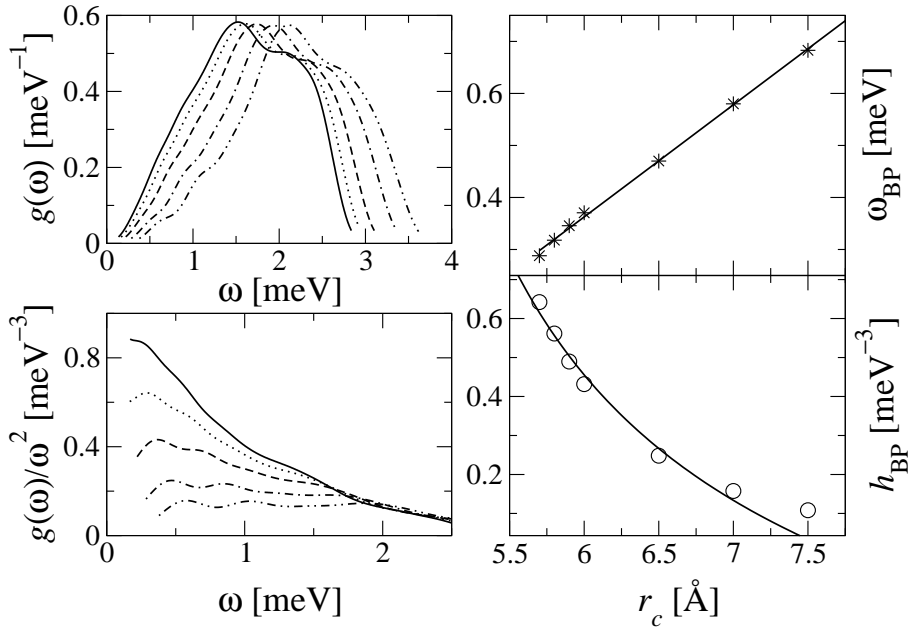}
\caption{The BP phenomenology for the PDZ binding domain. Left panels: DOS
  and DOS divided by the Debye law. Solid line $r_{c}=5.5$ \AA, dotted line
  $r_{c}=5.7$ \AA, dashed line $r_{c}=6$ \AA, dot-dashed line $r_{c}=6.5$
  \AA, dot-dot-dashed line $r_{c}=7$ \AA. Right panels: scaling of the BP
  frequency and height with the cutoff $r_{c}$. Solid lines represent the
  mean field prediction, $\omega_{BP}\sim (r_c-r_c^*)$, $h_{BP} \sim
  (r_c-r_c^*)^{-1/2}$. }
\label{fig:bp4panels}
\end{figure}

In Fig.~\ref{fig:bp4panels} we report our results for the vibrational DOS of
one particular protein out of the examined ensemble, namely the PDZ binding
domain. In order to obtain a smooth and more reliable DOS, we generate a
large number (order $10^3$) of surrogates whose topology is compatible with
the experimentally determined native structure~\cite{noiPRL}. The low
frequency region shows a strong dependence on $r_c$, since a non-trivial
excess of modes develops as a peak which eventually diverges for some
critical $r_c^*$. We identify this peak as the BP. The position and height
of the BP are plotted in the right panels versus the interaction cutoff
$r_c$. The mean field predictions are also shown to perform reasonably well
in the vicinity of the phase transition.

We have repeated this analysis on several proteins, choosing structures in a
wide range of sizes (from 51 to 578 residues), and we always found an
identical phenomenology.  In order to clarify the physical origin of this
behavior and its significance in biological terms, we investigate in the
next subsection the role of the critical interaction cutoff $r_c^*$ by
measuring its correlation with structural properties of the protein.

\subsection{The critical cutoff $r_c^{\scs \ast}$: correlation with 
  structural properties}

In Table~\ref{tab1} we report measures of selected geometrical and
structural indicators for a choice of proteins of different sizes, along
with the correlations with the corresponding value of the critical cutoff
distance $r_c^{\scs \ast}$.

\begin{itemize}

\item We find a positive correlation  between $r_c^{\scs \ast}$ and the
number of residues $N$.
       
\item We also find a positive correlation with a measure of the protein
volume $V$ estimated by constructing the equivalent ellipsoid according to
the procedure described in ref.~\cite{sheraga}. Such ellipsoid is
characterized by three principal radii $a_1$, $a_2$ and $a_3$, which are
obtained by diagonalising the symmetric matrix $\mathcal{A}_{\mu\nu} =
\sum_i r^\mu_i r^\nu_i /N $, where $\mu,\nu=1,2,3$ are Cartesian labels and
$r_i$ are the amino-acid coordinates. In the case of a continuous
distribution of mass inside the ellipsoid, it can be shown that the three
radii are simply given by $a_i = \sqrt{5 \lambda_i}$, where
$\lambda_1,\lambda_2, \lambda_3$ are the eigenvalues of the matrix ${\cal
A}$.

\item There clearly exists a negative correlation between $r_c^{\scs \ast}$ and the
packing coefficient.  The latter is a non-dimensional measure of compactness
defined as
\[
p = \frac{4 \pi}{3} \left( \frac{N}{V} \right) \left( \frac{d_{0}}{2} \right)^3
\]
where $d_{0} \simeq 3.83$ \AA \ is the distance between two consecutive
residues along the protein main chain.

\item The critical values $r_c^{\scs \ast}$ turn out to be positively
correlated with the fraction of flexible residues. Such measure is
calculated by flagging as highly fluctuating a residue whose average
fluctuation over the first three slow modes is above its overall average
fluctuation~\cite{demirel}. Average fluctuations over the first $n$ modes
are computed as
\[
\langle |\vec{u}_{i}|^2 \rangle = \frac{3 k_{B}T}{\kappa} \frac{1}{n}
\sum_{k=1}^{n} \omega_{k}^{-2} \xi_{i}^k \xi_{i}^k
\]
where $\vec{\xi}^k$ is the $k-$th eigenvector of the Hessian matrix
and $\omega^2_k$ the corresponding eigenvalue.

\item Finally, a poor correlation is found between $r_c^{\scs \ast}$ and a
combined measure of secondary structure content, defined as the fraction of
residues classified either as from $\alpha$-helices of
$\beta$-sheets~\cite{STRIDE}.

\end{itemize}
   
Overall, the measured correlations confirm that $r_c^{\scs \ast}$ is an
\emph{indirect measure of structural stability of a protein}, small values
of $r_c^{\scs \ast}$ corresponding to highly compact (higher packing
coefficient) and less flexible structures.  Moreover, $r_c^{\scs \ast}$ is
shown to increase with extensive quantities such as $N$ and $V$, which may
reflect the inherent increasing loss of globularity of larger structures.
Surprisingly enough, the degree of secondary structure content does not
prove to be an accurate indicator of stability in this sense. However, this
only indicates that $r_c^{\scs \ast}$ is mostly sensitive to the large-scale
degree of compactness, relevant to the robustness of the lowest vibrational
modes.  In other words, the local high stability of secondary motifs is not
equivalent to the large-scale one as probed by $r_c^{\scs \ast}$. This is
also evident from the poor correlation (-0.19) existing between the
secondary structure content and the packing coefficient $p$.

To be more clear, what we have found is that the excess of modes appearing
in the low frequency region of a protein's DOS is a precursory feature that
flags the approaching of a topological instability, whose meaning is to be
traced back to the analogy with glass-forming materials. In particular, as
it is the case for the Gaussian model in
glasses~\cite{grigera01,ciliberti03}, such excess of modes should be
interpreted as a precursor of the transition within a model that by
definition becomes meaningless at the critical point. In fact, as the
interaction cutoff $r_{c}$ is decreased below the typical range of the first
off-chain coordination shell, the model describes protein conformations that
start loosening until they eventually unfold, thus becoming closer and
closer to liquid-like assemblies of amino-acids.  From a biological point of
view, all that hints at an inherent effort of reconciling between spatial
properties of liquids, i.e. increased degree of mobility, and the necessity
of maintaining the structural stability of compact biological agents.

\begin{table}[t]
\tbl{Values of the critical cutoff distance $r_c^{\scs \ast}$ for
different proteins and correlation with geometrical and structural
indicators.}
{\begin{tabular}{@{}lccccccc}
\toprule
{\bf Protein} & PDB-id  & \quad $N$ \quad   &   $V$ (10$^4$ \AA$^3$) \quad  &  
\quad $p$ \quad &
 \% flex residues & $(\alpha+\beta)$-fract. & $r_c^\ast$ (\AA) \\
\hline
Insulin                      & \texttt{4INS}      &        51       &        0.75       &        0.20       &       27.45 &   0.53  &   4.57  \\
Protein G                    & \texttt{1PGB}      &        56       &        0.76       &        0.21       &       37.50 &   0.70  &   3.64  \\
Ubiquitin($^{\rm a}$)        & \texttt{1UBI}      &        71       &        1.04       &        0.20       &       35.21 &   0.46  &   3.53  \\
PDZ binding domain($^{\rm a}$) & \texttt{1BFE}      &        85       &        1.20       &        0.21       &       30.00 &   0.55  &   4.03  \\
Lysozyme                     & \texttt{166L}      &       162       &        2.75       &        0.17       &       43.21 &   0.74  &   4.27  \\
Adenylate Kinase             & \texttt{4AKE}      &       214       &        5.41       &        0.12       &       50.93 &   0.64  &   7.85  \\
LAO                          & \texttt{2LAO}      &       238       &        4.36       &        0.16       &       47.06 &   0.60  &   5.44  \\ 
CYSB                         & \texttt{1AL3}      &       260       &        4.40       &        0.17       &       41.15 &   0.59  &   4.70  \\
PBGD                         & \texttt{1PDA}      &       296       &        5.33       &        0.16       &       41.22 &   0.60  &   3.70  \\
Thermolysin                  & \texttt{5TLN}      &       316       &        5.02       &        0.18       &       40.51 &   0.53  &   4.55  \\
HSP70 ATP-binding domain     & \texttt{3HSC}      &       382       &        7.49       &        0.15       &       38.74 &   0.66  &   5.28  \\
Fab-fragment                 & \texttt{1AE6}      &       437       &        9.57       &        0.13       &       45.08 &   0.48  &   5.70  \\
Serum Albumin                & \texttt{1AO6}      &       578       &       14.55       &        0.12       &       47.06 &   0.70  &   5.70  \\\hline
Correlation with $r_c^{\scs \ast}$    &                    &      {\bf 0.45} &   {\bf 0.52}      &    {\bf  -0.82}   &    {\bf  0.67}  &  {\bf  0.17}&   1 \\
\botrule
\end{tabular}}
\tabnote{$^{\rm a}$ In these proteins short terminal tails of a few residues have been cut out in order to recover the correct low-frequency modes of the structure.}
\label{tab1}
\end{table}%

\subsection{Critical cutoff and functional motions: the case of PDZ binding domains}

The connection of the structural instability of a protein as revealed by the
Boson peak analysis and the details of its functional dynamics is a matter
of investigation deserving a systematic study on its own right. Here, we give a an example
of such analysis by sticking to the specific example of the PDZ binding domain.  

\begin{figure}[t!]
\begin{center}  
\includegraphics[width=12 truecm,clip]{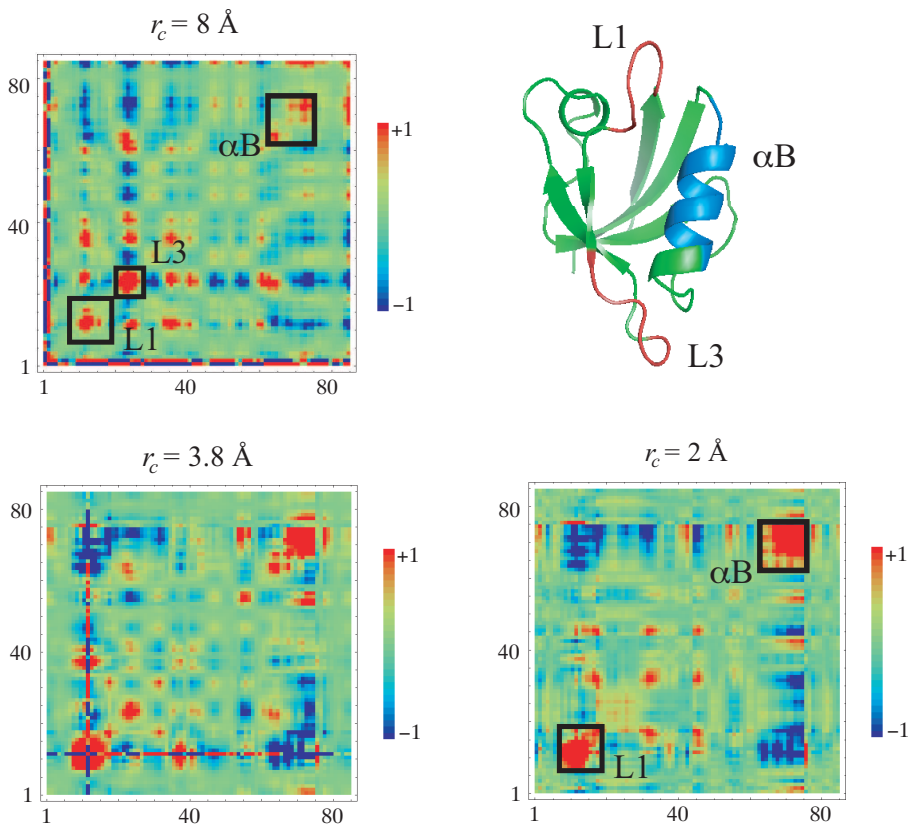}
  \caption{Pattern correlation plots of the second slowest normal mode as
           calculated through the Gaussian ANM model for the PDZ binding
           domain. The regions characterized by the largest displacements
           are highlighted and referenced explicitly within the protein
           structure. The functional motion corresponds to L1 and $\alpha$B
           moving out of phase (cleft opening/closing).}
  \label{f:PDZ}
\end{center}
\end{figure}

Postsynaptic density-95/disks large/zonula occludens-1 (PDZ) interaction
domains are crucial in regulating the cell dynamics. They play a fundamental
role in signaling pathways by organizing networks of receptors and in
targeting selected cellular proteins to multi-protein
complexes~\cite{PDZ1,PDZ2,PDZ3,PDZ4}.  Most PDZ-mediated interactions occur
through the recognition of C-terminal peptide motifs by a large binding
cleft built in the PDZ fold (see Fig.~\ref{f:PDZ}).  Interestingly,
experimental evidence from NMR measurements strongly suggests that the
dynamics of PDZ domains upon ligand binding show correlations over the
entire protein structure: this confirms the crucial role of collective
low-frequency modes in the biological functions of proteins~\cite{PDZ-NMR}.

In a previous work~\cite{PDZbpj}, a thorough normal-mode analysis of the PDZ
binding dynamics has been performed by means of the CHARMM all-atom force
field and of a recent version of the ANM, the C$\beta$-ANM~\cite{CbANM}. In
such a model, also C$\beta$ carbons are considered and the complex chemical
interactions between residues are described by springs connecting all
C$\alpha$-C$\alpha$, C$\alpha$-C$\beta$, and C$\beta$-C$\beta$ pairs whose
distances in the native fold are smaller than $r_{c} = 7.5$ \AA.  The main
result was that a single low-frequency mode captures the main spatial
correlations characterizing the binding dynamics, namely a concerted opening
of the hydrophobic pocket (see cartoon in Fig~\ref{f:PDZ}).

In order to investigate the correlation of such functional pattern with the
general phenomenon of structural instability occurring as the density is
lowered, we have looked for a mode with the same displacement template
within the Gaussian ANM model as a function of the parameter $r_{c}$.  We
observe that the known pattern is recovered in the second slowest mode, but
only at relatively low values of $r_{c}$, whereas the mode looses
cooperativity at larger values of the interaction cutoff. This is shown in
Fig.~\ref{f:PDZ}, where we report correlation plots for $r_{c}=2,3.8$ and
$8$ \AA. The correlation matrix $C$ for mode $k$ is defined as
\[
C_{ij}^k = \sum_{\alpha=x,y,z} \xi^{k}_{i\alpha} \xi^{k}_{j\alpha}
\]
where $\xi^{k}_{j\alpha}$ is the $j-$th component in directon $\alpha$ of
the $k-$th normalized eigenvector, so that $-1 \leq C_{ij} \leq 1$.

Our results indicate that, as $r_{c}$ is lowered towards the structural
instability, the mode pattern becomes more and more pronounced. In
particular, the loop L1 and the helix $\alpha$B become rigid,
anti-correlated units regulating the large-scale opening of the binding
cleft. At the same time, the displacement of loop L3 correlated to L1 for
tighter structures ($r_{c}$ = 8 \AA) gets strongly reduced. This result is a
clear signature of the high robustness of large-scale functional
motions. However, in order to assess the generality of our inferences, the
same analysis should be repeated on a large pool of protein structures with
known functional dynamics.

\section{Conclusions and perspectives \label{sect:conclu}}

In this paper we have addressed the Boson peak phenomenology experimentally
observed in proteins by means of a simple elastic network model. After
identifying the interaction cutoff as a control parameter, we have checked
numerically the analytical predictions of ERM theory, already successfully
reproduced by glass-forming materials. The BP turns out to be an intrinsic
feature of proteins spectra as it is for structural glasses. On the same
line, it signals the presence of a mechanical instability of the protein
structure which can be related to the unfolding transition. However, in
order to establish a precise relationship between BP divergence and
unfolding transition, a more systematic study based on a sharp-cutoff
elastic network model should be carried out.

To substantiate our interpretation of the BP emergence as related to the
presence of an underlying mechanical instability, we have measured the
correlation between the critical cutoff $r_c^*$ and structural properties of
the protein. The overall result indicates that $r_c^*$ may serve as an
indirect measure of global compactness of the protein, unrelated to the
presence of local more stable motifs, such as secondary structures. Another
fact to be stressed is that the critical $r_c$ is always smaller than $7\div
8$ \AA, which is supposed to be the ``correct'' value for $r_c$ obtained by
comparing elastic network models with molecular dynamics spectra. Moreover,
$r_c^*$ is always greater than, or at worst of the order of, the distance
between consecutive amino-acids along the backbone, ensuring that the loss
of global stability does not interfere with the polymeric nature of
proteins. As a rule of thumb, the larger $r_c^*$, the more unstable is the
protein.

Taken together with the experimental fact that proteins are typically active
in an environment whose local temperature is a few degrees smaller than the
unfolding one, all these observations lead to a new interpretation of the
surprising fact that proteins live more or less close to a phase
transition. According to our analysis, in order to be efficient molecular
machines able to perform their biological function, a protein has to keep a
relative mechanical rigidity while being able to easily access the local
minima directly connected to its native state. The BP is then the universal
signature of such a trade-off. Interestingly, the observation that
functional modes are those which ``survive'' in the critical regime strongly
supports this interpretation.

We think that, in order to establish a deeper connection between the
presence of the BP in proteins and this kind of theoretical approach,
more experimental work is needed. The supposed relationship between
the structural stability of a protein, its unfolding transition
temperature, and its vibrational features certainly deserves a more
detailed study.  Furthermore, a systematic experimental approach to
the BP phenomenology in proteins would allow to establish strength and
limits of the ERM theory with respect to real systems. For the sake of
concreteness, we would like to propose some ideas for experimental
investigation of the BP phenomenon in proteins that may prove useful
in probing the validity of our theoretical framework.

In general, high pressure is a useful tool for the study of protein
structure and dynamics~\cite{heremans}. More precisely, the effect
of pressure on proteins is two-fold. Moderate pressures (< 0.4 GPa)
cause elastic alterations in the spatial structure of the proteins,
while further increase in pressure can cause the loss of the secondary
structure along with the biological inactivation and denaturation. A
number of pressure-dependent spectroscopic measurements have revealed
non-trivial response of protein structures to pressure-induced stress,
such as secondary structure-specific volume fluctuations and structure
stabilization~\cite{akasaka}. Moreover, normal mode calculations
have also been successfully employed in such a context, showing that
contributions to volume fluctuations from low frequency normal modes
are typically found to dominate over those from higher frequency
modes~\cite{gopresNM}.

In view of the above facts, we propose to perform pressure-dependent
measurements in the low-to moderate pressure regime to investigate the
corresponding effect of density fluctuations on the BP frequency. In
particular, combining vibrational spectroscopy with global
experimental probes, such as adiabatic compressibility measurements
and small angle X-ray scattering, should\ enable one to investigate
experimentally the correlations between the critical interaction
cutoff $r_c^\ast$ and non-local structural indicators such as the
packing coefficient.

\emph{Acknowledgments} The work of S.~C. is supported by EC through the
network MTR 2002-00319, STIPCO.




\begin{thebibliography}{99}

\bibitem{green94}  J. L. Green, J. Fan, C. A. Angell,  
                   J. Phys. Chem. {\bf 98} 13780 (1994).

\bibitem{frauenfelder88} H Frauenfelder, F. Parak, R. D. Young,
                         Ann. Rev. Biophys. Biophys. Chem. {\bf 17} 451 (1988).

\bibitem{xie02}    A. Xie, L. Van Der Meer, R. H. Austin,
                   J. Biol. Phys. {\bf 28} 147 (2002).

\bibitem{piazza05} F. Piazza, P. De Los Rios, Y.-H. Sanejouand,
                   Phys. Rev. Lett. {\bf 94} 145502 (2005).

\bibitem{doster89} W. Doster, S. Cusack, W. Petry,  
                   Nature {\bf 337} 754 (1989).

\bibitem{iben89}   I. E. T. Iben, \emph{et al.}, 
                   Phys. Rev. Lett. {\bf 62} 1916 (1989).

\bibitem{orecchini01} A Orecchini, A Paciaroni, A. R. Bizzarri, S. Cannistraro, 
                      J. Phys. Chem. B {\bf 105} 12150 (2001). 

\bibitem{leyser99}  H. Leyser, W. Doster, M. Diehl,  
                    Phys. Rev. Lett. {\bf 82} 2987 (1999).

\bibitem{doster90}  W. Doster, S. Cusack, W. Petry,  
                    Phys. Rev. Lett. {\bf 65} 1080.

\bibitem{ansari85}  A. Ansari {\em et al.},  
                    Proc. Nat. Acad. Sci. {\bf 82} 5000 (1985).

\bibitem{onuchic97} J. N. Onuchic, Z. Luthey-Schulten, P. G. Wolynes,
                    Ann. Rev. Phys. Chem. {\bf 48} 545 (1997).

\bibitem{frauenfelder98} H. Frauenfelder, and D. T. Leeson, 	
                         Nature Struct. Biology {\bf  5} 757 (1998).

\bibitem{frauenfelder02} H. Frauenfelder,  
                         Proc. Nat. Acad. Sci.  {\bf 99} 2479 (2002).

\bibitem{buchenau92}     U. Buchenau, Yu. M. Galperin, V. L. Gurevich,
                         Phys. Rev. B {\bf 46} 2798 (1992)

\bibitem{karpov83}       V. G. Karpov {\em et al.}, 
                         Sov. Phys. JETP {\bf 57} 439 (1983).

\bibitem{schirmacher98}  W. Schirmacher, G. Diezemann, C. Ganter, 
                         Phys. Rev. Lett. {\bf 81} 136 (1998).

\bibitem{taraskin01}     S. N. Taraskin, Y. L. Loh, G. Natarajan, S. R. Elliott, 
                         Phys. Rev. Lett.  {\bf 86} 1255 (2001)

\bibitem{goetze00}       W. G\"otze, M. R. Mayr,  
                         Phys. Rev. E {\bf 61} 587 (2000). 

\bibitem{voigtmann02}    T. Voigtmann, 
                         J. Non-Cryst. Solids {\bf 307} 188 (2002).

\bibitem{grigera01}      T. S. Grigera, V. Mart\'\i{}n-Mayor, G. Parisi, P. Verrocchio, 
                         Phys. Rev. Lett. {\bf 87} 085502 (2001).

\bibitem{ciliberti03} S. Ciliberti, T. S. Grigera, V. Mart\'{\i}n-Mayor, 
  G. Parisi, P. Verrocchio, J. Chem. Phys. {\bf 119} 8577 (2003).

\bibitem{wyart05}  M. Wyart, S. R. Nagel, T. A. Witten,  Europhy. Lett. {\bf 72} 486 (2005).

\bibitem{noiPRL} S. Ciliberti, P. De Los Rios, F. Piazza, Phys. 
  Rev. Lett. {\bf 96} (198103) (2006).
  
\bibitem{Tirion}   M. M. Tirion, Phys. Rev. Lett. \textbf{77} 1905 (1996).

\bibitem{Bahar}    A.~R.~Atilgan  S.~R.~Durell,  R.~L.~Jernigan, M.~C.~Demirel, 
  O.~Keskin, I.~Bahar,     Biophys. J. {\bf 80} 505 (2001).
                     
\bibitem{Haliloglu} T. Haliloglu {\em et al.}, Phys. Rev. Lett.  \textbf{79}
                   3090 (1997).

\bibitem{hinsen}   K. Hinsen, Proteins \textbf{33} 417 (1998).

\bibitem{yhs}      F. Tama and Y. H. Sanejouand, Prot. Eng. \textbf{14} 1 (2001).

\bibitem{mezard99} M. M\'ezard, G. Parisi, A. Zee,  Nucl. Phys. {\bf B559} 689 (1999).

\bibitem{martin01} V. Mart\'{\i}n-Mayor, M. M\'ezard, G. Parisi,  P. Verrocchio, 
                   J. Chem. Phys {\bf 114} 8068 (2001).

\bibitem{grigera03} T. S. Grigera, V. Mart\'\i{}n-Mayor, G. Parisi,  P. Verrocchio,
                    Nature {\bf 422} 289 (2003).

\bibitem{sheraga}   M. H. Hao, {\em et al.},  Proc. Nat. Acad. Sci. {\bf 89} 6614 (1992).
  
\bibitem{demirel}   M. C. Demirel, O. Keskin,  J. Biomol. Struct. Dyn. {\bf 22} 381 (2005).
                        
\bibitem{STRIDE}    D. Frishman, P. Argos, Proteins {\bf 23}(4) 566 (1995).

\bibitem{R3}        J. R. Banavar, A. Flammini, D. Marenduzzo, A. Maritan,  A. Trovato,  
                    Complexus {\bf 1} 4 (2003).
                        
\bibitem{PDZbpj}    P. De Los Rios, {\em et al.}, Biophys. J. {\bf 89}  14 (2005).  

\bibitem{PDZ1}      A. Y. Hung, M. Sheng, J. Biol. Chem. {\bf 277} 5699 (2002).

\bibitem{PDZ2}      M. Sheng,  C. Sala, Annu. Rev. Neurosci. {\bf 24} 1 (2001).

\bibitem{PDZ3}      M. Zhang, W. Wang,  {\bf 36}Ê530 (2003).

\bibitem{PDZ4}      T. Pawson, P. Nash, Science. {\bf 300} 445 (2003).

\bibitem{PDZ-NMR}   E. J. Fuentes, C. J. Der, A. L. Lee,  J. Mol. Biol. {\bf 335} 1105 (2004).

\bibitem{CbANM}     C. Micheletti, P. Carloni, A. Maritan,  Proteins {\bf 55} 635 (2004).
                    
\bibitem{heremans} K. Heremans, L. Smeller, \emph{Biochim. Biophys Acta}, 
  \textbf{1386}, 353-370 (1998).

\bibitem{akasaka}  K Akasaka, H Li, H Yamada, R Li, T Thoresen
 and CK Woodward, {\em Protein Science}, \textbf{8}(10) 1946-1953 (1999).

\bibitem{gopresNM} T. Yamato, J. Higo, Y. Seno and N. Go, 
{\em Proteins Struct. Funct. Genet.}, \textbf{16}, 327-340 (1993). 

\end{thebibliography}
\end{document}